\documentclass[11pt]{article}
\usepackage{fullpage}
\usepackage{graphicx}           
\usepackage{bm}                 
\usepackage{amsmath}            
\usepackage{amsfonts}
\usepackage{amssymb}
\usepackage{latexsym}           

\usepackage{natbib}
\bibpunct[,]{[}{]}{,}{n}{,}{,}

\newcommand{\md}{\ensuremath{\mathrm{d}}}

\newcommand{\ket}[1]{\mbox{$|#1\rangle$}}
\newcommand{\bra}[1]{\mbox{$\langle#1|$}}

\newcommand{\identity}{\leavevmode\hbox{\small1\kern-3.2pt\normalsize1}}

\begin{document}

\title{Where to quantum walk}
\author{%
Viv Kendon\thanks{Contact: \texttt{V.Kendon@leeds.ac.uk}}\\
{\textit{\small School of Physics and Astronomy, University of Leeds, Woodhouse Lane, Leeds, LS2 9JT, UK.}}\\
{\textit{\small NORDITA, Roslagstullsbacken 23, 106 91 Stockholm, Sweden.}}\\
}%

\date{31st May 2011}

\maketitle

\begin{abstract}
Quantum versions of random walks have diverse applications that are
motivating experimental implementations as well as theoretical studies.
However, the main impetus behind this interest is their use in quantum
algorithms, which have always employed the quantum walk in the form of a
program running on a quantum computer.  Recent results showing that quantum
walks are ``universal for quantum computation'' relate entirely to algorithms,
and do not imply that a physical quantum walk could provide a new
architecture for quantum computers.  Nonetheless, quantum walks used
to model transport phenomena in spin chains and biomolecules broaden
their scope well beyond algorithms, and reopen the question of when a physical
implementation might provide useful computational outputs.  In this article
we determine the conditions under which a physical quantum walk experiment
could provide useful results beyond the reach of classical computation.
\end{abstract} 


\section{Introduction} 
\label{sec:intro}

Quantum versions of random walks were introduced in the late 80s
by \citeauthor{gudder88} \cite{gudder88}
and \citeauthor{aharonov92a} \cite{aharonov92a}.
Early work by \citeauthor{grossing88a} \cite{grossing88a}
and \citeauthor{meyer96a} \cite{meyer96a,meyer96b,meyer96c}
focused on the more general case of quantum cellular automata.
The first application to quantum algorithms was by
\citeauthor{farhi98a}~\cite{farhi98a} using a continuous time quantum walk.
\citeauthor{aharonov00a}~\cite{aharonov00a} and
\citeauthor{ambainis01a}~\cite{ambainis01a} described
discrete time quantum walks from an algorithmic perspective, starting
a decade of increasingly intense study of their applications and properties.
In the simplest setting of a line or a Cartesian lattice,
on which the walker hops between integer sites,
a quantum walk spreads quadratically faster than a classical random walk.
This provides the speed up in many quantum walk algorithms,
e.g., for basic searching as presented by
\citeauthor{shenvi02a}~\cite{shenvi02a}.
For a recent review of quantum walk searching
see \citeauthor{santha08a}~\cite{santha08a}.
%
Exponential speed up has been proved for crossing a hypercube by
\citeauthor{kempe02a}~\cite{kempe02a,kempe02c} 
more efficiently than a classical random walk, and
\citeauthor{childs02a}~\cite{childs02a} proved a 
continuous time quantum walk can find its way through a particular
``glued trees'' graph exponentially faster than any classical algorithm.
On the other hand, the phenomenon of Anderson localization \cite{anderson58a}
of quantum particles is well-known, and has been related to continuous time
quantum walks by \citeauthor{keating06a}~\cite{keating06a}.
Studies by \citeauthor{krovi06a}~\cite{krovi05a,krovi06a,krovi07a}
confirm this extends to the discrete time setting, and highlight the
importance of symmetry in the behaviour of quantum walks.
\citeauthor{stefanik07a} \cite{stefanik07a,stefanik08a}
have characterized the localization of unbiased quantum walks on
regular lattices, showing how it depends on the topology of the lattice,
the coin operator and chosen initial state.

Quantum walks provide useful models of physical phenomena,
such as quantum state transfer in spin chains as introduced by
\citeauthor{bose03a} \cite{bose03a} and
reviewed by \citeauthor{kay09a} \cite{kay09a},
or energy transport in biomolecules (\citeauthor{mohseni08a}
\cite{mohseni08a}).
Perfect quantum state transfer
can be achieved for particular configurations
(reviewed in \citeauthor{kendon10b} \cite{kendon10b}),
of interest for building quantum wires in quantum computers.
Highly efficient transport can be obtained by using
imperfect quantum walks where the amount of decoherence
is tuned to optimise the quantum walk properties
\cite{mohseni11b,kendon02c}.

Experimental implementations of quantum walks were proposed from
the beginning \cite{aharonov92a},
the first photonic realisation \cite{bouwmeester99a}
was recognised retrospectively by \citeauthor{sanders02a} \cite{sanders02a}.
Implementation using atoms trapped in an optical lattice
as proposed by \citeauthor{dur02a} \cite{dur02a} was recently
performed by \citeauthor{karski09a} \cite{karski09a}.  Their atoms maintained
full quantum coherence for up to ten steps, with decoherence
degrading the walk to classical spreading after 25 steps.
A quantum walk on four sites, as a computation in a three qubit
NMR quantum computer, was implemented by \citeauthor{ryan05a} \cite{ryan05a},
who performed state tomography on all eight steps of the periodic dynamics
and also applied decoherence to demonstrate the transition to a classical
random walk.
Recent work by \citeauthor{schreiber11a} \cite{schreiber11a} achieved
28 fully coherent steps of a photonic quantum walk, mapping the 
position to time delays in the photon propagation provided by fibre loops.
Using time delays to encode the locations allows a single set
of components to be used for all sites and steps of the quantum walk;
the detection system records the time of arrival of the photons
with sufficient accuracy to distinguish all the
encoded locations.
They also demonstrated 11-step walks with fully controlled decoherence
of three different types, static disorder leading to Anderson
localisation, fast fluctuations returning the walk to classical
behaviour, and slow fluctuations that produce even faster spreading
when averaged over many walks.
Photonic quantum walks can also be performed in substrates containing
the sites.
\citeauthor{broome10a} \cite{broome10a} used separate calcite crystals for each
step of the walk, with the position mapped to the displacement of the walker
across the crystal.
Coupled photonic waveguides for quantum walks were first
employed by \citeauthor{perets07a} \cite{perets07a}
performing a continuous time quantum walk across 100 lattice sites.
Multiple correlated photons can easily be injected into such devices
\cite{peruzzo10a}.
However, these waveguides contain a fresh set of sites for each step of 
the walk, a significant extra resource compared to designs that reuse the
same components.

Recent work by \citeauthor{childs09a}~\cite{childs09a} proving that 
quantum walks are ``universal for quantum computation'' has renewed
interest in their role in quantum information.  However, it is important
to direct research efforts in the most effective directions.  This paper
places this important result in context, and explores the computational
potential of quantum walk experiments.  A brief overview of quantum 
walks is given in the next section, followed in section \ref{sec:algo}
by a description of how quantum walks are used as a tool
for quantum algorithms.  In section \ref{sec:compare} the limitations of
classical simulations of quantum walks are described, and the possibilities
for usefully expanding beyond these limits with quantum walk experiments are
explored.  Conclusions are summarised in section \ref{sec:conc}.

\section{Quantum Walks}
\label{sec:qwalk}

Like classical random walks, quantum walks have both discrete time
\cite{aharonov92a,watrous98a,aharonov00a,ambainis01a},
and continuous time \cite{farhi98a} versions.  
We briefly review the simplest walk on a line, since many experiments
have focused on this instance.
A discrete quantum walk on a line is defined in direct analogy with
a classical random walk, where the walker steps one unit forward
or back based on the outcome of a random coin toss.
The quantum walker has a quantum coin that will in 
general be in a superposition, so the quantum walker
ends up in a superposition of positions on the line.
However, the coin toss is not random, since pure quantum dynamics need
to be unitary, but if measured, the outcome of the measurement
would be random, like the classical coin.
We define a two-dimensional quantum coin with basis states 
$\ket{{\pm 1}}$ and label the position basis states $\ket{x}$
by the integer location on the line $x$. 
The evolution of the walk is governed by a coin
operator that acts on the quantum coin at each step of the walk.
The simplest choice is the Hadamard coin operator,
\begin{equation}
\mathbf{C}_2^{(\text{Had})}=\frac{1}{\sqrt{2}}\left( \begin{array}{cc}
        1 & 1\\
        1 & -1
        \end{array} \right).
\label{eq:had}
\end{equation}
Varying the phase $\beta$ in the initial coin state $\ket{c_0}$:
\begin{equation}
\ket{c_0} = \sqrt{b}\ket{{-1}} + e^{i\beta}\sqrt{1-b}\ket{{+1}},
\end{equation}
where $b$ is the bias in the initial state, and $\beta$ is the
relative phase between the two components, produces outcomes
with varying skewness \cite{bach02a}.
After ``tossing'' the coin with the coin operator, the particle moves to 
adjacent positions conditioned on the coin state
\begin{eqnarray}
&&\mathbf{S}\ket{{-1},x} = \ket{{-1},x-1}\nonumber\\
&&\mathbf{S}\ket{{+1},x} = \ket{{+1},x+1} .
\label{eq:cshift}
\end{eqnarray}
Here we have written $\ket{c,x}\equiv\ket{c}\otimes\ket{x}$ for
combined basis states.
One step of the quantum walk is produced by the unitary operator
$\mathbf{U} = \mathbf{S}(\mathbf{C}\otimes\identity_x)$.
A quantum walk of $T$ steps starting from an initial state
$\ket{\psi_0}$ is thus 
\begin{equation}
\ket{\psi_T} = \mathbf{U}^T\ket{\psi_0}.
\label{eq:psiT}
\end{equation}

\begin{figure}[tbh!]
  \begin{minipage}{1.0\columnwidth}
    \begin{center}
        \resizebox{0.55\columnwidth}{!}{\rotatebox{0}{\includegraphics{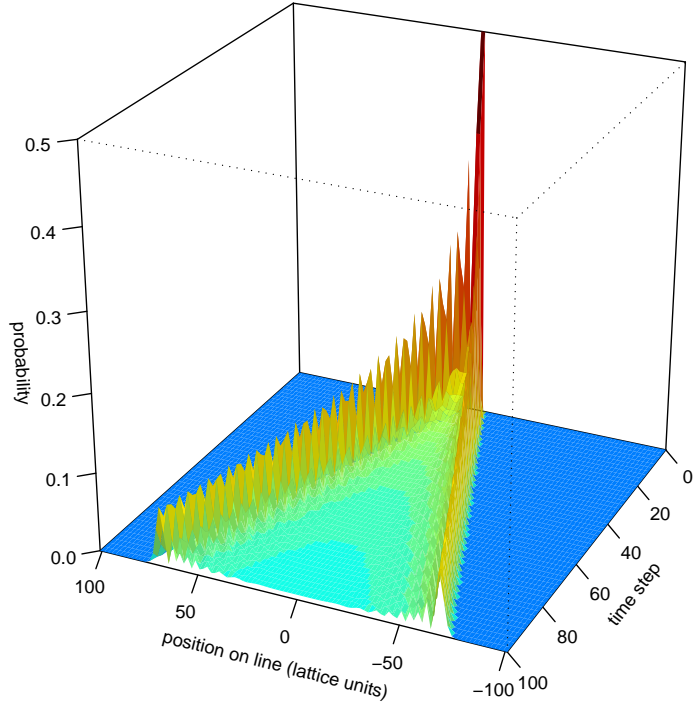}}}
    \end{center}
  \end{minipage}
  \caption{\textit{Probability distribution for a quantum walk on the line
        over 100 steps showing evolution of system, using a Hadamard coin,
        equation (\ref{eq:had}),
        and a symmetric initial state $(\ket{{+1},0}+i\ket{{-1},0})/\sqrt{2}$.
        Only positions with non-zero probability of occupation are shown,
        since odd positions are unoccupied at even time steps and vice versa.}}
  \label{fig:walk1d}
\end{figure}

The position probability distribution of a quantum walk on a line
is by now well-known, the time evolution over 100 steps with a
Hadamard coin operator and initial state of 
$\frac{1}{\sqrt{2}}(\ket{{+1,0}}+i\ket{{-1,0}})$ 
is shown in figure \ref{fig:walk1d}.
The double-peaked spreading expands at a linear rate, giving a
quadratic speed up over the $\sqrt{T}$ spread of the binomial distribution
produced by a classical random walk on the line.

The continuous time quantum walk on a line is even simpler. The quantum 
walker has a transition probability $\gamma$ per unit time to hop left 
or right to neighbouring positions.  This can be expressed as a Hamiltonian
\begin{equation}
\mathbf{H} = \sum_{x=-\infty}^{x=+\infty} \gamma\left\{\ket{x-1}\bra{x} + \ket{x+1}\bra{x}\right\}.
\end{equation}
The quantum walk evolves according to the Schr\"odinger equation,
\begin{equation}
i\frac{\md}{\md t}\ket{\psi(t)} = \mathbf{H}\ket{\psi(t)},
\end{equation}
with formal solution
\begin{equation}
\ket{\psi(t)} = \exp\{-i\mathbf{H}t\}\ket{\psi(0)},
\label{eq:psit}
\end{equation}
where $\ket{\psi(t)} = \sum_x a_x(t)\ket{x}$ and the $\{a_x(t)\}$ are complex
numbers with normalisation $\sum_x|a_x(t)|^2 = 1$.
The time evolution is very similar
to the discrete time quantum walk on the line depicted
in figure \ref{fig:walk1d}, with the same quadratic speed up over
its classical counterpart.

Discrete and continuous quantum walks have been solved analytically on
regular structures and their asymptotic scaling described in detail
\cite{ambainis01a,aharonov00a,konno02b}.
Differences in their behaviour compared with classical random walks
are also found when there are absorbing boundaries \cite{bach02a}.
Both discrete and continuous time quantum walks generalise straightforwardly
from the line to arbitrary connected structures (graphs), usually described
by their adjacency matrix $\mathbf{A}$.  An edge between sites $x$ and $y$
is represented by the $xy$th entry in $\mathbf{A}$ being one, with zeros elsewhere.
For the continuous time quantum walk, the Hamiltonian becomes $\mathbf{H}=\gamma \mathbf{A}$.
The discrete time walk needs a mapping between the coin states and the choice of
edge at each sites \cite{kendon04a,kendon03c}.

\section{Algorithmic quantum walks}
\label{sec:algo}

Algorithmic uses of quantum walks are always
binary encoded in the quantum computer.
Without this, the efficiency gains they provide would be negated.
Binary encoded means that the position of the walker
is stored as a superposition of binary numbers in a qubit register.
For simplicity, we will illustrate this for the discrete time quantum walk
on the line, but it is easily generalised to any labeled set of positions.
The state of the quantum walker after $T$ steps,
\begin{equation}
\ket{\psi_T} = U^T\ket{\phi_0} = \sum_{c,x} a_{c,x}(T)\ket{c,x},
\label{eq:amps}
\end{equation}
with complex amplitudes $a_{c,x}(T)$ satisfying $\sum_{c,x}|a_{c,x}(T)|^2 = 1$,
can equally well be interpreted as the superposition state of
a quantum register where $x$ written in binary is the pattern of
zeros and ones, and $c$ is an extra bit recording the state of the coin.
For a quantum walk of $T$ steps, the furthest it can travel is
to $x = \pm T$, so we need $\lceil\log_2(2T+1)\rceil$ qubits,
plus one more for the coin, in our quantum register.
A quantum walk of a million steps thus requires just 22 qubits.
The continuous time quantum walk needs to be discretised for quantum
computation and this can be done efficiently,
see \citeauthor{childs02a} \cite{childs02a}, using
similar resources to a discrete time walk \cite{kendon03c}.

\subsection{Universality of quantum walks}
\label{ssec:universal}

\begin{figure}[t!]
  \begin{center}
     \begin{minipage}{0.2\columnwidth}
	\flushright
	\Large
	$\ket{\psi_1}_{\text{in}}$

	\vspace{2em}

	$\ket{\psi_2}_{\text{in}}$
     \end{minipage}
     \begin{minipage}{0.55\columnwidth}
        \resizebox{0.95\columnwidth}{!}{\rotatebox{0}{\includegraphics{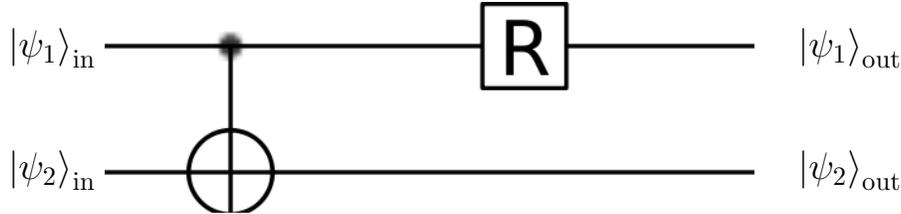}}}
     \end{minipage}
     \begin{minipage}{0.2\columnwidth}
	\Large
	$\ket{\psi_1}_{\text{out}}$

	\vspace{2em}

	$\ket{\psi_2}_{\text{out}}$
      \end{minipage}
  \end{center}
  \caption{\textit{Simple quantum circuit with two qubits labeled
$\ket{\psi_1}, \ket{\psi_2}$, where 
the dot and circle connecting the two qubits is a controlled-\textrm{NOT} operation,
and} \textsf{R} \textit{is a rotation through $\pi/8$.}}
  \label{fig:qcircuit}
\end{figure}

\begin{figure}[tbh!]
  \begin{center}
     \begin{minipage}{0.2\columnwidth}
	\flushright
	\large
	$\ket{00}_{\text{in}}$

	\vspace{2.4em}

	$\ket{01}_{\text{in}}$

	\vspace{2.4em}

	$\ket{10}_{\text{in}}$

	\vspace{2.4em}

	$\ket{11}_{\text{in}}$

	\vspace{3ex}

     \end{minipage}
     \begin{minipage}{0.55\columnwidth}
        \resizebox{0.95\columnwidth}{!}{\rotatebox{0}{\includegraphics{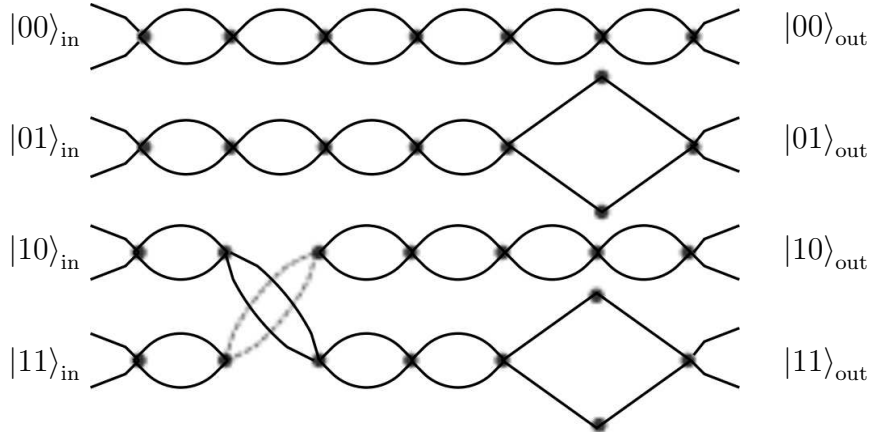}}}
     \end{minipage}
     \begin{minipage}{0.2\columnwidth}
	\large
	$\ket{00}_{\text{out}}$

	\vspace{2.4em}

	$\ket{01}_{\text{out}}$

	\vspace{2.4em}

	$\ket{10}_{\text{out}}$

	\vspace{2.4em}

	$\ket{11}_{\text{out}}$

	\vspace{1ex}

	\null
      \end{minipage}
  \end{center}
  \caption{\textit{Quantum walk graph corresponding to figure \ref{fig:qcircuit}.
	The dotted edges pass underneath the solid edges without joining.
	For details, see} \citeauthor{lovett09b} \cite{lovett09b}.}
  \label{fig:circuitgraph}
\end{figure}

Recent work by \citeauthor{childs09a} \cite{childs09a}
proving that ``quantum walks are universal for quantum computation''
uses the correspondence between Hamiltonians and continuous time quantum walks
to prove that any quantum algorithm can be recast as a quantum walk
algorithm.  To ensure the algorithm is efficient, the Hamiltonian
must be ``suitably sparse'' \cite{childs02a,aharonov03a,berry05a},
as explained below.
The discrete time quantum walk has also been shown to
be universal in the same sense \cite{lovett09b},
using a similar mapping between
the circuit model of quantum computing and a quantum walk on a graph
derived from the circuit.
A simple example is illustrated in
figures \ref{fig:qcircuit} and
\ref{fig:circuitgraph}.
Comparison of these figures shows clearly that the graph on which the
quantum walk takes place is exponentially larger than the quantum circuit.
%

A fully general Hamiltonian over $n$ qubits has $2^n\times 2^n$ terms,
requiring an exponential amount of information to describe it.
It therefore cannot represent an efficient quantum computation.
To obtain an efficient quantum computation,
the graph on which the quantum walk takes place must have a description
that is logarithmic in the number of sites -- this corresponds to it
being ``suitably sparse''.  This is clearly the case for the graph in
figure \ref{fig:circuitgraph}, since it has been derived from the circuit
in figure \ref{fig:qcircuit}.
If the description of the classical structure of sites
on which the quantum walk takes place is inconveniently large, the job of
storing and accessing the description is given to an oracle, as in \citeauthor{childs02a} \cite{childs02a}.
Such oracles are a theoretical computer science concept, not 
intended to be implemented in an actual computation or experiment.
The proof by \citeauthor{childs09a} that quantum walks are universal for
quantum computation allows the mathematics of quantum walks to be applied
more generally to theoretical computer science, potentially facilitating
advances in both areas.
It does not suggest that a physical quantum walk on a structure like that of
figure \ref{fig:circuitgraph} is an efficient way to build a 
general purpose quantum computer.

\section{Comparing classical and quantum}
\label{sec:compare}

Although algorithmic uses of quantum walks do not require
physical quantum walk experiments,
this still leaves the possibility that a quantum walk experiment
could provide useful computation of quantum walks themselves.
At first sight this appears to blur the distinction between
computation and experiment.  We will avoid circularity (and philosophical
issues) by taking a computation in this context to mean starting from a mathematical
model -- in this case the quantum walk dynamics as described by equations 
(\ref{eq:psiT}) or (\ref{eq:psit}) -- and calculating properties
of it.  A quantum walk experiment that accomplishes this must thus be
set up to match the particular quantum walk under study.
This requires well-characterised experimental apparatus 
to provide confidence it is really doing the correct quantum walk.
We will, of course, test our experiment by performing simple quantum walks for
which the outcomes are already known.  In other words, we do an experiment
to check our apparatus matches the mathematical description we have for it,
then we do the experiment to perform the quantum walk computation we
really want.  The extrapolation is not guaranteed to work, but similar
uncertainty is inherent in numerical simulation.
The computer code used to do a classical computation of a quantum walk
may have errors, and similar tests using known quantum walks
are done to reduce this possibility.

The reason why a quantum walk experiment could be efficient enough
to be computationally useful is because it avoids much of the overhead
created by translating the quantum walk dynamics into numbers and
elementary gate operations in a computer, be it quantum or classical.
This can only give us an advantage
for particular cases, because the exponential saving in memory by
encoding the position as a binary number in a quantum computer
will always win eventually as the quantum walk gets larger and longer.  
But we don't yet have large quantum computers, the most powerful general
purpose computational devices we have are classical computers.
So we need to compare quantum walk experiments with current classical
computational capabilities.

\subsection{Classical computational limits}
\label{ssec:classical}

Classical numerical simulation of a quantum walk
is a straightforward evaluation of equation (\ref{eq:psiT}) or
(\ref{eq:psit}), or their generalisations for different graphs.
In high-level computational environments, such as Matlab, it takes only
a few lines of code.  As with any full simulation of a quantum system,
we need to store and manipulate all of the complex amplitudes $a_{c,x}(t)$
describing the wavefunction, as given in equation (\ref{eq:amps}).
Each amplitude requires two floating point numbers (real and imaginary parts).
Using 32 bits (4 bytes) for each floating point number, we can
store $2^{27}$ amplitudes in 1 Gbyte of memory.  For comparison, a new desktop
computer in 2011 probably has around 4 Gbyte of memory in total.
As already noted, a quantum walk on the line of a million steps
needs only 22 qubits, described by $2^{22}$ amplitudes, which requires
about 4Mbyte of memory, and this computation is
quick and easy on a desktop computer.
One of the largest simulations of a quantum system to date was not of a quantum walk, but of 36 qubits, requiring $2^{36}$ amplitudes.
\citeauthor{deraedt06a} \cite{deraedt06a} employed many computers
in parallel with a combined total of 1 Terabyte of memory
-- a new desktop computer in 2011 probably has a hard disk of this size.
Each additional qubit doubles the memory required,
so it is easy to appreciate that a quantum computer of just
40 logical qubits could outperform any classical computer.

\subsection{Experiments and scaling}
\label{ssec:expt}

However, quantum walks are not qubit systems, a physical quantum
walk on a line of a million steps (easily computed on any desktop computer)
covers about two million sites,
which is a significant experimental challenge, whether
the sites are spatial as in \citeauthor{karski09a} \cite{karski09a}
or mapped to time delays as in \citeauthor{schreiber11a} \cite{schreiber11a}.  
It is not possible to save resources by having a ``quantum
substrate'' that stores the sites in some sort of superposition.
This would amount to having a ``quantum hash table'',
which was proved impossible by \citeauthor{nayak99a} \cite{nayak99a}.
The quantum coherence required for a million steps is even more challenging
than building the substrate,
but higher dimensional walks cover the ground faster.
For a two-dimensional lattice, two million sites is $1414\times 1414$,
and supports only 706 time steps.  In three dimensions,
two million sites is roughly $125^3$, taking just 62 steps to fill.
This is not far from the number of steps in
recent experimental quantum walks on a line.
The big difference is that you need a six-dimensional coin for a walk
on a 3D lattice, although this is not a problem in principle,
see \citeauthor{hamilton11a} \cite{hamilton11a} for a proposal
utilising orbital angular momentum of photons.
The most recent experiments have achieved quantum walks on tens
\cite{karski09a,schreiber11a}, or at most
a hundred sites \cite{perets07a},
so there is some way to go before the classical
computational capabilities of millions of sites can be beaten.

\subsection{Random substrates}
\label{ssec:random}

Quantum walks on regular structures are easier to implement
because the symmetry simplifies various aspects of the task.
These symmetries also simplify the classical computation, and in
many cases facilitate analytic solution too.
A more difficult problem
is quantum walks on disordered systems, such as percolation
lattices.  A recent numerical study in \citeauthor{leung10a}
\cite{leung10a}, of 1D and 2D percolation lattices probed the
average behaviour over many random percolation lattices.
On a line they studied 10,000 steps, about  $2^{14}$ locations,
with around 5000 repetitions over different percolations.
In two dimensions, they ran simulations for $140$ time steps,
$281\times 281$, which is roughly $2^{16}$ lattice sites.
On a workstation, it took a week to do 5000 repetitions
with different percolation lattices.
The task that took most time was actually generating the
pseudo-random numbers needed to create each percolation lattice.
To carry out a similar calculation with a quantum walk experiment
requires a way to produce random percolation lattices
for the quantum walk to take place on.  Generating the randomness
is not difficult in itself, any laboratory has sufficiently
random sources of noise.  Arranging for the randomness to create a
percolation lattice for the quantum walk requires a new experimental design.

\subsection{Strong versus weak simulation}
\label{ssec:weaksim}

There is a further issue that needs to be considered when comparing
quantum and classical computations of quantum walks.  The classical
computation as described in this article gives as output the values
of all of the complex amplitudes $a_{c,x}(T)$ defining the quantum state.
This is known as ``strong simulation'' \cite{VandenNest2009}.
Experimental quantum walks provide just one output, usually
a location at which the quantum walker ended up.
Further information can only be obtained by performing
many repetitions of the experiment, to build up the probability
distribution.  Classical simulation that is
strictly equivalent to this is known as ``weak simulation''.  It
uses less resources per run, but similarly requires many repetitions
to obtain more information.  With clever algorithmic or experimental
design, it may be possible to perform a final measurement
that yields more useful information directly (the mean or standard
deviation, perhaps).  With photonic quantum walks in particular,
it may be so easy to perform many repetitions that this issue does
not constitute a problem, even for producing sufficient statistics to
construct the whole probability distribution.

\subsection{Decoherence}
\label{sec:decoher}

As well as perfect quantum walks, it is also important to
understand what happens when imperfections affect the quantum walk
dynamics.  While some simple models of Markovian noise can be
solved analytically, many cases require numerical simulation \cite{kendon06b}.
Using models in which the noise is provided by nonlinear
operators in a standard master equation, the quantum state 
is now represented by a density matrix, and the
classical simulations require roughly the square of the
resources necessary for a pure quantum state.
This reduces the size of the largest practical classical
simulation of a quantum walk down to
around $2^{12}$ sites on a workstation.
This is only $64^2$ or $16^3$, so the number of steps, even in
two dimensions, is comparable with current experiments.
In a classical computation, individual control of the decoherence
at each site and step can be done easily, the extra resources being
proportional to the size of the description of the required controls.
In an experiment, arbitrary controls add significant complexity
to the design, see \citeauthor{schreiber11a} \cite{schreiber11a} for example.

\subsection{Multiple walkers}
\label{sec:multi}

There is one important case where a physical implementation can win 
even sooner over classical computers:
when there are multiple interacting walkers.
Multiple non-interacting walkers only add particle statistics
(for indistinguishable walkers), they live in a predictable
subspace of the full Hilbert space that can
be simulated as efficiently with classical computers as single walkers.
Analytical and numerical study of non-interacting
multiple walkers are not as advanced as
for single walkers, but many cases of interest are analytically tractable,
\cite{mayer10a,omar06a}.
However, when an interaction is added between walkers occupying the same site,
the resources required for classical simulation grow quickly.
With $m$ walkers on $L$ locations, the full Hilbert space is of size $L^m$.  
The exponential growth in the size of the Hilbert space is the
same as for qubits, hence the size of such classical simulations is
limited to $L^m < 2^{40}$ for current capabilities.
Multiple interacting quantum walkers have been studied from the earliest
days of quantum walks: they are quantum cellular automata.
They have been proved capable of universal quantum computation,
for a review of their significance and properties,
see \cite{wiesner08a}.  Quantum cellular automata are
particular suited to optical lattice experiments. 
A scheme with two coupled walkers has been
proposed by \citeauthor{albrecht11a} \cite{albrecht11a}, as 
a first step in this direction.

\section{Conclusions}
\label{sec:conc}

A Hilbert space of size $2^{40}$ corresponding to 40 spin-1/2
systems is beyond current classical capabilities.
Quantum walks with a single walker and a Hilbert space of this size
can be performed by a qubit quantum computer with 40 qubits, but
a physical quantum walk of this size would require around $10^{12}$ sites
to be provided, a more challenging experiment than a 40 qubit quantum computer.
Ion trap and optical lattice experiments already trap many more than
40 ions and atoms, but not yet with the dynamical control required
for general operations.

When making the comparisons it should be remembered that
classical simulations can provide information
about the complete quantum wavefunction, whereas a single run of
a quantum walk experiment provides only one measurement outcome.
Repetitions of the experiment thus need to be quick and simple
if a single result doesn't provide a useful answer.
This favours photonic implementations over trapped atoms or ions.

Decohering quantum walks are more challenging for a classical
computation, reducing the largest classical simulation size
to around $10^5$ sites.
However, while classical simulations of this size can be fully
general without significant extra overhead, the experimental
challenges may limit the types of decoherent quantum walks that can be
successfully created in the laboratory.

Although single quantum walker experiments will find it hard to beat
classical computers, 
multiple interacting quantum walkers (quantum cellular automata)
are a realistic goal for experiments that reach beyond the scope
of classical numerical simulation.  They have a Hilbert space of
size $L^m$ for $m$ walkers on $L$ sites.  For $L=10$ sites, $m=12$ walkers
could outperform a classical computer.
The optimal architecture for this is atoms or ions in optical lattices,
since the natural interactions between the trapped atoms or ions are
well-suited to the task.
The immediate challenge is to identify quantum walk-based problems
worth solving -- in any architecture -- that may be experimentally
accessible before general purpose quantum computers are built, and
to design the experiments that can implement them.

\subsection*{Acknowledgments}

Thanks to Erika Andersson, Christine Silberhorn and Andrew White for
useful discussions.
Inspiration for the title came from my friend Reb Gowers'
novel ``When to Walk''.
VK is funded by a UK Royal Society University Research Fellowship.  



\bibliography{../bibs/qrw,../bibs/misc,../bibs/qbio,../bibs/qit,../bibs/KLBbibextra}



\end{document}